\documentclass{article}
\usepackage{array}
\newcolumntype{L}{>{\centering\arraybackslash}m{3cm}}
\newcolumntype{M}[1]{>{\centering\arraybackslash}m{#1}}
\usepackage{amsmath}

\usepackage[utf8]{inputenc} 
\usepackage[T1]{fontenc}    
\usepackage{hyperref}       
\usepackage{url}            
\usepackage{booktabs}       
\usepackage{amsfonts}       
\usepackage{nicefrac}       
\usepackage{microtype}      
\usepackage{graphicx}
\usepackage{xcolor}
\usepackage{xfrac}
\usepackage[font=small,labelfont=bf,tableposition=top]{caption}
\usepackage[font=footnotesize]{subcaption}
\usepackage{wrapfig}
\usepackage{adjustbox}
\usepackage{multirow}
\usepackage{algorithm}
\usepackage{algorithmic}
\usepackage{tablefootnote}

\definecolor{zzttqq}{rgb}{0.48,0.70,0.33}
\definecolor{aabbcc}{rgb}{0.63,0.77,0.90}
\newcommand{\removelatexerror}{\let\@latex@error\@gobble}

\title{Edge, Fog, and Cloud Computing : An Overview on Challenges and Applications\\}

\author{%
  Thong Vo, Pranjal Dave, Gaurav Bajpai, and Rasha Kashef \\
   Electrical, Computer, and Biomedical Engineering\\
  Ryerson University
}

\begin{document}

\maketitle
\begin{abstract}
With the rapid growth of the Internet of Things (IoT) and a wide range of mobile devices, the conventional cloud computing paradigm faces significant challenges (high latency, bandwidth cost, etc.). Motivated by those constraints and concerns for the future of the IoT, modern architectures are gearing toward distributing the cloud computational resources to remote locations where most end-devices are located. Edge and fog computing are considered as the key enablers for applications where centralized cloud-based solutions are not suitable. In this paper, we review the high-level definition of edge, fog, cloud computing, and their configurations in various IoT scenarios. We further discuss their interactions and collaborations in many applications such as cloud offloading, smart cities, health care, and smart agriculture. Though there are still challenges in the development of such distributed systems, early research to tackle those limitations have also surfaced. 
\end{abstract}

\section{Introduction}

Different service offerings like Infrastructure-as-a-Service, Platform-as-a-Service, and  Software-as-a-Service have arisen as forms of the cloud technology. These service types provide great services at lower costs. Cloud computing had the capability to manage Big Data and hence it was accepted as a cost-effective solution. The Internet of Things (IoT) comprises billions of devices with smart sensors and computing power connected with each other via various networking protocols. These sensors generated Big Data and were available at lower prices. The use of sensors led to numerous solutions in different applications in agriculture, industry, healthcare and smart cities. As new use-cases arrive, the computing resources in the cloud face a problem in storing and processing the raw data into useful information. An example of this is video streaming where the quality of video keeps on increasing making it difficult for a centralized system to handle everything. Instead, we can process the data at the end-user. This is an example of the recent advances applied in the cloud computing paradigm. Edge Computing and Fog Computing which are associated with Edge Analytics are  used in many IoT applications where there is a middle layer which is called the Fog Layer between the cloud and edge for processing, storage, and computation. Edge Computing has various differences from the conventional Cloud Computing systems. In this paper, we analyze the reason for which we need Edge and Fog computing. Next, we understand both computing technologies and survey their advantages. We also discuss several applications which have been developed to solve existing problems using edge computing methodologies. Furthermore, we look at the potential challenges we might face when we develop a system using Edge Computing. \\

The remainder of this paper is described as follows. The survey starts in Section II with definitions of Edge and Fog Computing. In Section III, we discuss various applications and systems developed to solve current problems. Section IV covers the probable reasons due to which these systems might face certain challenges. Finally, we conclude the paper in Section V.

\section{Edge Computing and Fog Computing}

The edge and fog computing paradigm proves to be more efficient to solve local complex analytical tasks.  
        
\subsection{Edge computing}
Edge computing provides low latency and proved to be effective, thus leading to a new area in research where computations were done towards the edge of the network. In other terms, edge computing is a computational paradigm in which computation occurs close to the source of created data. This process prevents the present issues in the cloud-like storage and the low system throughput. 

\subsection{Fog computing}
 Fog computing brings new options of computing in addition to the edge and the cloud. Generally, this infrastructure in the center is decentralized. The name comes from the thought that fog, which seems like the cloud, is close to the surface, indicating that this paradigm brings infrastructure close to the end devices. In a practical scenario, a cloud would contain thousands of computing resources whereas a fog layer can contain millions of decentralized nodes. We need scaling of these many resources as we assume that we have billions of smart things collecting data using embedded sensors.Each layer in the Fog Computing paradigm is connected to another layer via gateways. We see that as we get near the end-user, the number of devices increases. The latency also increases as we got closer to the core or the cloud infrastructure. 
 
\section{Applications}
\subsection{Cloud Offloading}
The addition of fog nodes can significantly reduce the communication latency between applications and the centralized cloud, which in turn improves the Quality of Service. Those computational nodes can be located close to IoT devices to offload information to the cloud. The authors in \cite{Yousefpour2018OnRI} introduce a general framework and offloading policies to assess and minimize delay in the communication process between IoT, fog, and the cloud. The proposed policies consider comprehensive interactions between entities in the networks such as from IoT to fog, from fog to cloud, and from edge to edge communication schemes. In addition, the policies take into account both the number of pending commands and the types of those commands. The study also develops an analytical model for evaluating delays in general IoT, fog, cloud frameworks that are applicable for a wide variety of applications. Fog computing also enables heavy computational tasks to be transferred to base stations or IoT gateways when computing power onboard mobile robots or drones is limited. For such resources to be effectively utilized, it is necessary to distribute computational jobs to entities in the network such that critical latencies are retained at acceptable rates. The authors in \cite{Kattepur2017APrioriEO} estimate the actual offloading times using profiling tools that leverage execution time on development testbeds. They verified their proposals with image editing applications, video processing tasks and robotic mission plans. Precog \cite{Drolia2017PrecogPF} leverages selective processing on the devices to limit the requirement for picture uploading to the cloud in order to lower the recognition delay of cloud-based image recognition services. The edge nodes proactively cache classification results for future predictions. Initially, sets of features of the input image are extracted, and these feature sets are classified into related content. In the future, if a similar request/image feature is presented, the cache can fulfill that request. Otherwise, a cache miss occurs, and the related feature set is sent to the cloud for recognition. The authors also propose a prefetching mechanism to load recognition information into the cache before the actual request arrives to further reduce cache miss. They demonstrate that their system can reduce the delay up to five times, utilize resources efficiently, and improve recognition accuracy.

\subsection{Data Stream Processing}
While various streaming applications exist, they are typically deployed in the centralized cloud. In \cite{Cheng2015GeeLyticsGE}, they proposed a streaming analytics solution for distributed environments where data comes from different regions, and analytical latency is sensitive to users and applications. To fulfill these requirements, they propose an edge analytics system to dynamically process data at the edge and the cloud in real-time. Their system supports many stream types such as media streams and event streams at a large scale with up to one thousand connected components. The set-up targets to produce quick analytics results yet maintain a low communication bandwidth to the cloud. Additionally, they discuss a strategy to deploy the application in a distributed environment with the pub/sub mechanism. Along with minimal latency, another concern for streaming video analytics is the demanding computational resource to generate real-time outputs. The authors in \cite{Ananthanarayanan2017RealTimeVA} present a traffic video analytics system that not only yields high quality analytical results but also maintains its resource consumption at a low level. Live video feeds from crossroads in a city are analyzed continuously to identify dangerous traffic patterns and estimate levels of traffic. The platform includes public cloud options, private edge nodes with various hardware to process media information, track objects, and execute algorithms. The systems access current required resources for current requests and assign optimal computing entities for the tasks. Despite being built and verified in one city, the platform is applicable to many other cities around the world.

\subsection{Smart City}
With the increase in globalization comes many problems such as traffic jams, an increase in energy consumed and waste from industries and the heat island effect. The urban population must deal with these issues and their negative impacts.  To have an environmental balance and reduce metropolitan challenges, smart-city solutions have been the focal point over the past two decades. Smart cities are now a reality with the advancement of the technological tools needed to support them. Transitioning to a smart technology-supported city is now possible with Internet of Things (IoT) devices, edge-fog-cloud computing, and 5G telecommunication networks, big data, artificial intelligence, and geographical information. Smart sensors are being widely used to improve transportation, planning utility and parking, monitoring pollution, and health care. Smart City technologies provide near real-time knowledge of problems requiring attention. To make a city’s functioning optimal, the dynamic data from devices/sensors is collected, processed, analyzed at the edge, fog or the cloud computing layer to provide a time-sensitive solution to real-time applications/operations. Bhardwaj et al. \cite{Bhardwaj2019IoTES} proposed two setups to find the differences in the time for response and the bandwidth use in Cloud and Fog. They installed IoT sensors in their university campus. Using fog computing processing time, the number of hops traversed and bandwidth usage dropped. Sapienza et al. \cite{Sapienza2016SolvingCE} talk about a mobile edge model that boosts the speed of data transfer and communication in smart cities. Their approach gives computing power to network devices that were earlier only used for communication. This would be handy for smart cities from being held back by latency issues. Sigwele et al. \cite{Sigwele2018IntelligentAE} have suggested a 5G-based smartphone gateway exclusively for healthcare in a smart city. Their approach is partitioning and loading the tasks for execution on virtual computers. Their study asserted that the proposed protocol improved energy efficiency and lowered service response time.
Santos et al. \cite{Santos2018FogCE} use 5G network communication with fog technology framework. It is based on the fact that the nodes in the fog layer communicate with cloud using the Open-Shortest-Path-First rule. The bandwidth utilization and latency in fogs are reduced with respect to cloud computing.

\subsection{Smart Agriculture}
Smart Agriculture incorporates modern technology in agriculture and livestock. The aim is to increase profits in terms of revenue and cause less damage to the environment. Sensors mounted on farm machines such as tractors and trucks and in fields and farms aid in the collection of real-time data directly about planting, spraying, produce, in-season imagery, soil types, and weather. An important goal of smart agriculture is to analyze large volumes of this data for informed decision-making by applying all modern technology in big data and computing. IoT and Edge computing can make farms more beneficial and viable. The different sensors installed on the farm to gather real-time data and monitor the environment with Edge Computing and Cloud for data robustness. Chen et al. \cite{Chen2021IdentificationOF} study overcome the transmission of images upon detecting pests to cloud by combining edge computing and GPU - a neural network which is stored on NVIDIA Jetson TX2 and a drone that photographs the pest. The Tiny-YOLOv3 integrates Jetson and Raspberry Pi with edge intelligence for object recognition. The TX2 transfers to the cloud layer the farm coordinates and then uses a computer or mobile device to analyze the data. The agricultural drone system helps farmers act in real-time to identify the exact pest location which decreases the use of pesticide and causes less damage to the environment and increase productivity and profits. Liu et al. \cite{Liu2021DesignAI} study the crops and soil condition using a real-time monitoring system that uses solar energy for powering various sensors. The system collects humidity of the air, temperature and salinity of the soil. The edge information is stored in a local memory card and then through the 4G network is transmitted to the cloud server. They report that the system has been unattended and maintenance free for nearly two years. The correlation between soil-salinity and growth can be analyzed to produce quality agricultural products.Sakthi et al. \cite{Sakthi2020SmartAK} proposed a system to help the farmers make intelligent choices about farming. The five-layered system consists of sensors in the agricultural environment layer, edge layer collects data and executes machine learning algorithms, fog layer servers execute clustering algorithm and reduce data volume transfer to the cloud, cloud layer executes knowledge discovery pattern generation and an interface layer using web-pages and smart-phone. Tsipis et al. \cite{Tsipis2020LatencyAdjustableCC} propose a network which has the various stakeholders and sensor devices that generate the data, Wireless Sensor Networks (WSNs) in the edge layer, the fog layer devices forming a computing network with storage and servers, and cloud infrastructure with large number of storage and servers. The edge layer is realized using Arduino, Raspbian, ZigBee hardware and software platforms. They find in their study that the intermediate fog computing network reduces the average response time and offers the farmers the opportunity for almost real-time monitoring.

\subsection{Healthcare}
In this section we will cover applications of edge and fog computing in the healthcare domain. Edge computing and Fog computing provide solutions that have low latency and also satisfy needs for energy consumption. Many smart devices like smartwatches and smart monitors collect data in real time. The existing state-of-the-art machine learning techniques can be applied to this data to get meaningful insights of the patient and get accurate predictions. M. Hartmann et al. \cite{Hartmann2019EdgeCI} discussed edge computing solutions in healthcare. They also discussed various papers that proposed new edge computing techniques and their contributions to the medical field. The authors also discussed various performance indicators like transmission and retrieval, encryption and authentication. In the end, the authors talked about future research challenges. Kaur et al. \cite{Kaur2017HealthMB} proposed a system to measure vital body parameters. The system is an example of the use of IoT. It uses Raspberry Pi sensors. Remote health monitoring is possible in this case and the system is also wearable. The authors used the KG011 heart rate sensor and DS18B20 temperature sensor. The authors also used the Node-RED software and the Message Queueing Telemetry Transport (MQTT) protocol. The combination of softwares and sensors provided an excellent accuracy cost trade off for their system. Klonoff et al. \cite{Klonoff2017FogCA}  addressed the issue of data processing for diabetes patients’ data. Moreover, the authors focused on general advantages of edge and fog computing in the healthcare field. For diabetes monitoring, the authors discussed sensors like continuous insulin pumps and glucose monitors. All the sensors are connected to a fog layer which maintains processing of the data and the processed data is stored on the cloud. The authors mentioned Dexcom G5 Platinum sensor and Medtronic Enlite system sensor to measure diabetes data of patients. The paper also mentions several devices used in other chronic medical conditions like cardiac defibrillators, cardiac pacemakers, and closed loop ventilator systems.

\subsection{Other Applications}
Table \ref{tab:apps} describes additional industrial applications of edge, fog and cloud computing in various sectors.

\begin{table}[t]
\caption{Applications of Edge and Fog Computing}
\label{tab:apps}
\small
\begin{tabular}{|M{1.6cm}|M{3cm}|M{3cm}|M{3cm}|M{0cm}}
\cline{1-4}
Study & Applications & Objectives & Devices involve &  \\ \cline{1-4}
Syafruddin et al. \cite{Syafrudin2018AnAF} & Smart Factories & Detect abnormal events on Assembly line & Humidity, Light and, Gas Sensor, Single board computer (SBC) &  \\ \cline{1-4}
Yar et al. \cite{Yar2021TowardsSH} & Smart Home & Fast Motion Detection, Reduce Energy consumption & Infrared Sensors,  Raspberry Pi &  \\ \cline{1-4}
Korupalli et al. \cite{Kumar2020InternetOT} & Smart Vehicles & Fog Computing in Transport Systems & Body Sensor Network &  \\ \cline{1-4}
Razfar et al. \cite{razfar2022weed} & Smart Agriculture &  Edge Computing in Agriculture & Infrared Sensors,  Raspberry Pi &  \\ \cline{1-4}
Karmari et al. \cite{karami2020smart} & Smart Transportation &  Computing in Transport Systems &  Sensor Network &  \\ \cline{1-4}
Jebamikyous et al. \cite{jebamikyous2022autonomous}\cite{jebamikyous2021deep} & Smart Transportation &  Edge Computing in Transportation & Camera-based &  \\ \cline{1-4}
Aljasim et al. \cite{aljasim2022e2dr} & Smart Transportation  &  Edge Computing in Transportation & Raspberry Pi and Camera-based &  \\ \cline{1-4}
Ghasemieh et al. \cite{ghasemieh20223d} & Smart Transportation  &  Edge Computing in Transportation & Sensors and Camera-based &  \\ \cline{1-4}
Sharma et al. \cite{sharma2021face} & Smart Healthcare  &  Edge Computing in Healthcare & Raspberry Pi and Camera-based &  \\ \cline{1-4}
Razfar et al. \cite{razfar2021assessing}\cite{razfar2021comprehensive} & Smart Healthcare  &  Edge Computing in Healthcare & Wearable Sensors &  \\ \cline{1-4}
Manjunath et al. \cite{manjunath2021distributed} & Smart e-commerce Systems &  Computing in e-commerce & Distributed Network &  \\ \cline{1-4}
\end{tabular}
\end{table}

\section{Challenges}
In this section , we present some drawbacks of our surveyed applications. In addition, we provide some early research efforts to tackle such limitations.

\subsection{Security}
Because IoT endpoints and related computing components are decentralized by design, it could be difficult to ensure sufficient security protection for their applications. Side-channel attacks can stem from the communication channels that connect IoT components, and some common problems are spoofing attacks and denial-of-service attacks. These challenges pose considerable threats to the development and operation of large-scale IoT systems. Selective restriction of access policies in the network can be used to minimize the impacts when threats are present in the network, monitoring network access to determine normal behavior and detect anomalies to contain harmful threats, and standardized communication protocols and design standards can reduce the efforts to maintain large scale systems are potential solutions to the IoT security challenges \cite{Sill2017StandardsAT}.

\subsection{Multiple-objective system design}
The majority of the surveyed applications only consider a few targets (e.g., latency, resource optimization, quality of service) and presume other indicators remain intact. In fact, to ensure satisfactory results for such applications, holistic indicators have to be considered before deploying into the real world. In order to tackle such challenges, it is desirable to design schemes that consider multiple objectives at once \cite{Haddadpajouh2020AI4SAFEIoTAA}. The new designs can provide some flexibility for end-users to choose preferable indicators to optimize or change depending on their needs. Such systems are more practical to be deployed in the real world where various indicators have their own set of requirements.

\section{Conclusions}
With the rapid development of systems that generate data that had high volume, velocity, and variety, there was a need to change the way in which real-time dynamic data was handled. The present cloud computing solutions were not fast enough and research came up with a new idea called Edge Computing. In this paper, we provided definitions of edge computing and fog computing and mentioned their differences. Edge computing and Fog computing presented solutions in several scenarios and made the use of dynamic data possible.
In this paper, we also discussed and surveyed various applications of Edge, Fog and Cloud Computing in the IoT world. We covered different research areas like Cloud Offloading and Data Stream processing, Smart Cities, Smart grids, Smart Agriculture and Healthcare. The use of the Edge computing paradigm in these areas proved very effective and there were several benefits of it over conventional cloud computing systems. We also discussed various challenges faced by Edge Computing systems. We conclude that the edge computing paradigm proves to be extremely beneficial. Its advantages definitely outweigh its disadvantages. Furthermore, as we move ahead in time, the rise in data generated will always be present in exponential amounts. Also, the use of IoT devices is expected to increase in the future. This is the reason why the use of edge and fog computing paradigms will prove efficient and effective. 

\bibliographystyle{ieeetr}
\bibliography{wvSent_acl2011}

\begin{thebibliography}{10}

\bibitem{Yousefpour2018OnRI}
A.~Yousefpour, G.~Ishigaki, R.~Gour, and J.~P. Jue, ``On reducing iot service
  delay via fog offloading,'' {\em IEEE Internet of Things Journal}, vol.~5,
  pp.~998--1010, 2018.

\bibitem{Kattepur2017APrioriEO}
A.~Kattepur, H.~K. Rath, and A.~Simha, ``A-priori estimation of computation
  times in fog networked robotics,'' {\em 2017 IEEE International Conference on
  Edge Computing (EDGE)}, pp.~9--16, 2017.

\bibitem{Drolia2017PrecogPF}
U.~Drolia, K.~Guo, and P.~Narasimhan, ``Precog: prefetching for image
  recognition applications at the edge,'' {\em Proceedings of the Second
  ACM/IEEE Symposium on Edge Computing}, 2017.

\bibitem{Cheng2015GeeLyticsGE}
B.~Cheng, A.~Papageorgiou, F.~Cirillo, and E.~Kovacs, ``Geelytics:
  Geo-distributed edge analytics for large scale iot systems based on dynamic
  topology,'' {\em 2015 IEEE 2nd World Forum on Internet of Things (WF-IoT)},
  pp.~565--570, 2015.

\bibitem{Ananthanarayanan2017RealTimeVA}
G.~Ananthanarayanan, P.~Bahl, P.~Bod{\'i}k, K.~Chintalapudi, M.~Philipose,
  L.~Ravindranath, and S.~Sinha, ``Real-time video analytics: The killer app
  for edge computing,'' {\em Computer}, vol.~50, pp.~58--67, 2017.

\bibitem{Bhardwaj2019IoTES}
A.~Bhardwaj and S.~Goundar, ``Iot enabled smart fog computing for vehicular
  traffic control,'' {\em EAI Endorsed Transactions on Internet of Things},
  2019.

\bibitem{Sapienza2016SolvingCE}
M.~Sapienza, E.~Guardo, M.~Cavallo, G.~L. Torre, G.~Leombruno, and
  O.~Tomarchio, ``Solving critical events through mobile edge computing: An
  approach for smart cities,'' {\em 2016 IEEE International Conference on Smart
  Computing (SMARTCOMP)}, pp.~1--5, 2016.

\bibitem{Sigwele2018IntelligentAE}
T.~Sigwele, Y.-F. Hu, M.~Ali, J.~Hou, M.~Susanto, and H.~Fitriawan,
  ``Intelligent and energy efficient mobile smartphone gateway for healthcare
  smart devices based on 5g,'' {\em 2018 IEEE Global Communications Conference
  (GLOBECOM)}, pp.~1--7, 2018.

\bibitem{Santos2018FogCE}
J.~Santos, T.~Wauters, B.~Volckaert, and F.~D. Turck, ``Fog computing: Enabling
  the management and orchestration of smart city applications in 5g networks,''
  {\em Entropy}, vol.~20, 2018.

\bibitem{Chen2021IdentificationOF}
C.-J. Chen, Y.-Y. Huang, Y.-S. Li, Y.-C. Chen, C.-Y. Chang, and Y.-M. Huang,
  ``Identification of fruit tree pests with deep learning on embedded drone to
  achieve accurate pesticide spraying,'' {\em IEEE Access}, vol.~9,
  pp.~21986--21997, 2021.

\bibitem{Liu2021DesignAI}
Y.~Liu, Y.~Wang, S.~Xu, W.~Hu, and Y.~jing Wu, ``Design and implementation of
  online monitoring system for soil salinity and alkalinity in yangtze river
  delta tideland,'' {\em 2021 IEEE International Conference on Artificial
  Intelligence and Industrial Design (AIID)}, pp.~521--526, 2021.

\bibitem{Sakthi2020SmartAK}
U.~Sakthi and J.~D. Rose, ``Smart agricultural knowledge discovery system using
  iot technology and fog computing,'' {\em 2020 Third International Conference
  on Smart Systems and Inventive Technology (ICSSIT)}, pp.~48--53, 2020.

\bibitem{Tsipis2020LatencyAdjustableCC}
A.~Tsipis, A.~Papamichail, G.~Koufoudakis, G.~Tsoumanis, S.~E. Polykalas, and
  K.~Oikonomou, ``Latency-adjustable cloud/fog computing architecture for
  time-sensitive environmental monitoring in olive groves,'' 2020.

\bibitem{Hartmann2019EdgeCI}
M.~Hartmann, U.~S. Hashmi, and A.~Imran, ``Edge computing in smart health care
  systems: Review, challenges, and research directions,'' {\em Transactions on
  Emerging Telecommunications Technologies}, 2019.

\bibitem{Kaur2017HealthMB}
A.~Kaur and A.~Jasuja, ``Health monitoring based on iot using raspberry pi,''
  {\em 2017 International Conference on Computing, Communication and Automation
  (ICCCA)}, pp.~1335--1340, 2017.

\bibitem{Klonoff2017FogCA}
D.~C. Klonoff, ``Fog computing and edge computing architectures for processing
  data from diabetes devices connected to the medical internet of things,''
  {\em Journal of Diabetes Science and Technology}, vol.~11, pp.~647 -- 652,
  2017.

\bibitem{Syafrudin2018AnAF}
M.~Syafrudin, N.~L. Fitriyani, G.~Alfian, and J.~Rhee, ``An affordable fast
  early warning system for edge computing in assembly line,'' {\em Applied
  Sciences}, 2018.

\bibitem{Yar2021TowardsSH}
H.~Yar, A.~S. Imran, Z.~A. Khan, M.~Sajjad, and Z.~Kastrati, ``Towards smart
  home automation using iot-enabled edge-computing paradigm,'' {\em Sensors
  (Basel, Switzerland)}, vol.~21, 2021.

\bibitem{Kumar2020InternetOT}
K.~V.~R. Kumar, K.~D. Kumar, R.~K. Poluru, S.~M. Basha, and M.~P.~K. Reddy,
  ``Internet of things and fog computing applications in intelligent
  transportation systems,'' 2020.

\bibitem{razfar2022weed}
N.~Razfar, J.~True, R.~Bassiouny, V.~Venkatesh, and R.~Kashef, ``Weed detection
  in soybean crops using custom lightweight deep learning models,'' {\em
  Journal of Agriculture and Food Research}, vol.~8, p.~100308, 2022.

\bibitem{karami2020smart}
Z.~Karami and R.~Kashef, ``Smart transportation planning: Data, models, and
  algorithms,'' {\em Transportation Engineering}, vol.~2, p.~100013, 2020.

\bibitem{jebamikyous2022autonomous}
H.-H. Jebamikyous and R.~Kashef, ``Autonomous vehicles perception (avp) using
  deep learning: Modeling, assessment, and challenges,'' {\em IEEE Access},
  vol.~10, pp.~10523--10535, 2022.

\bibitem{jebamikyous2021deep}
H.-H. Jebamikyous and R.~Kashef, ``Deep learning-based semantic segmentation in
  autonomous driving,'' in {\em 2021 IEEE 23rd Int Conf on High Performance
  Computing \& Communications; 7th Int Conf on Data Science \& Systems; 19th
  Int Conf on Smart City; 7th Int Conf on Dependability in Sensor, Cloud \& Big
  Data Systems \& Application (HPCC/DSS/SmartCity/DependSys)}, pp.~1367--1373,
  IEEE, 2021.

\bibitem{aljasim2022e2dr}
M.~Aljasim and R.~Kashef, ``E2dr: a deep learning ensemble-based driver
  distraction detection with recommendations model,'' {\em Sensors}, vol.~22,
  no.~5, p.~1858, 2022.

\bibitem{ghasemieh20223d}
A.~Ghasemieh and R.~Kashef, ``3d object detection for autonomous driving:
  Methods, models, sensors, data, and challenges,'' {\em Transportation
  Engineering}, vol.~8, p.~100115, 2022.

\bibitem{sharma2021face}
H.~Sharma, H.~Sewani, R.~Garg, and R.~Kashef, ``Face mask detection: A
  real-time android application based on deep learning modeling,'' in {\em 2021
  IEEE 12th Annual Information Technology, Electronics and Mobile Communication
  Conference (IEMCON)}, pp.~0106--0112, IEEE, 2021.

\bibitem{razfar2021assessing}
N.~Razfar, R.~Kashef, and F.~Mohammadi, ``Assessing stroke patients movements
  using inertial measurements through the advances of ensemble learning
  technology,'' in {\em 2021 IEEE 23rd Int Conf on High Performance Computing
  \& Communications; 7th Int Conf on Data Science \& Systems; 19th Int Conf on
  Smart City; 7th Int Conf on Dependability in Sensor, Cloud \& Big Data
  Systems \& Application (HPCC/DSS/SmartCity/DependSys)}, pp.~1482--1489, IEEE,
  2021.

\bibitem{razfar2021comprehensive}
N.~Razfar, R.~Kashef, and F.~Mohammadi, ``A comprehensive overview on iot-based
  smart stroke rehabilitation using the advances of wearable technology,'' in
  {\em 2021 IEEE 23rd Int Conf on High Performance Computing \& Communications;
  7th Int Conf on Data Science \& Systems; 19th Int Conf on Smart City; 7th Int
  Conf on Dependability in Sensor, Cloud \& Big Data Systems \& Application
  (HPCC/DSS/SmartCity/DependSys)}, pp.~1359--1366, IEEE, 2021.

\bibitem{manjunath2021distributed}
Y.~S.~K. Manjunath and R.~F. Kashef, ``Distributed clustering using multi-tier
  hierarchical overlay super-peer peer-to-peer network architecture for
  efficient customer segmentation,'' {\em Electronic Commerce Research and
  Applications}, vol.~47, p.~101040, 2021.

\bibitem{Sill2017StandardsAT}
A.~Sill, ``Standards at the edge of the cloud,'' {\em IEEE Cloud Computing},
  vol.~4, pp.~63--67, 2017.

\bibitem{Haddadpajouh2020AI4SAFEIoTAA}
H.~Haddadpajouh, R.~Khayami, A.~Dehghantanha, K.~R. Choo, and R.~M. Parizi,
  ``Ai4safe-iot: an ai-powered secure architecture for edge layer of internet
  of things,'' {\em Neural Computing and Applications}, pp.~1--15, 2020.

\end{thebibliography}

\end{document}